\documentclass[aps,prx,twocolumn,superscriptaddress]{revtex4-2}
\usepackage{amssymb}
\usepackage{amsfonts}
\usepackage{amsmath}	
\usepackage{pxfonts}
\usepackage{graphicx}
\usepackage{eucal}
\usepackage{mathrsfs}
\usepackage{theorem}
\usepackage{pifont}
\usepackage{color}
\usepackage{newtxmath}
\usepackage{multirow}
\usepackage{xcolor}
\usepackage{hyperref}

\newcommand{\figref}[1]{Fig.~\ref{#1}}

\newcommand{\rin}{r_\text{in}}
\newcommand{\rout}{r_\text{out}}

\begin{document}

\author{Chandana Mondal}
\affiliation{Dept. of Chemical Physics, The Weizmann Institute of Science, Rehovot 76100, Israel}
\author{Michael Moshe} 
\affiliation{Racah Institute of Physics, The Hebrew University of Jerusalem, Jerusalem, Israel 9190}
\email{michael.moshe@mail.huji.ac.il}
\author{Itamar Procaccia}
\affiliation{Dept. of Chemical Physics, The Weizmann Institute of Science, Rehovot 76100, Israel, Center for OPTical IMagery Analysis and Learning, Northwestern Polytechnical University, Xi'an, 710072 China}
\author{Saikat Roy}
\affiliation{Department of Chemical Engineering, Indian Institute of Technology Ropar, Punjab 140001, India}
\author{Jin Shang}
\affiliation{School of Physics and Astronomy, Shanghai Jiao Tong University, 200240 Shanghai, China}
\author{Jie Zhang}
\affiliation{Institute of Natural Sciences and School of Physics and Astronomy, Shanghai Jiao Tong University, 200240 Shanghai, China}





\title{Experimental and Numerical Verification of Anomalous Screening Theory in Granular Matter}

\begin{abstract}
	The concept of mechanical screening is widely applied in solid-state systems.
	Examples include nucleation of defects in crystalline materials, scars and pleats in curved crystals, wrinkles in strongly confined thin sheets, and cell-rearrangements in biological tissue. 
	Available theories of such screening usually contain a crucial ingredient, which is the existence of an ordered reference state, with respect to which screening elements nucleate to release stresses.
	In contradistinction, amorphous materials in which a unique reference state does not exist, nevertheless exhibits plastic events that act as screening geometric charges with significant implications on the mechanical response. 
	In a recent paper [Phys. Rev. E {\bf 104}, 024904] it was proposed that mechanical strains in amorphous solids can be either weakly or strongly screened by the formation of low or high density of plastic events. 
	At low densities the screening effect is reminiscent of the role of dipoles in dielectrics, in only renormalizing the elastic moduli. 
	The effect of high density screening has no immediate electrostatic analog and is expected to change qualitatively the mechanical response, as seen for example in the displacement field.
	On the basis of experiments and simulations, we show that in granular matter, strong screening results in significant deviation from 
	elasticity theory.
	The theoretical analysis, which  accounts for an emergent inherent length scale,
	the experimental measurements and the numerical simulations of frictional granular amorphous assemblies are in agreement with each other, and provide a strong support for the novel continuum theory.  
\end{abstract}


\maketitle


Granular matter exhibits properties that distinguish it from classical solids and liquids \cite{de1999granular}.
In particular, its macroscopic mechanics mixes elastic and plastic responses \cite{10KLP}, in addition to exotic flow properties \cite{jaeger1996granular}. Theoretically, it requires concepts and ideas that go beyond the classical  theories of solids and fluids. Viscoelasticity is an example of an accepted theory that combines fluid- and solid-like features, yet it fails in describing the mechanics of granular matter . A noteworthy example is the sudden dynamical arrest of a granular matter flowing through a funnel \cite{tang2011granular}, reflecting the celebrated pressure-controlled jamming transition associated with granular matter \cite{liu1998jamming}. An adequate macroscopic theory of granular matters that encompasses the rich possible responses to external stresses and strains
is still currently unavailable, posing a fundamental challenge for the study of such materials.

One major difficulty in developing a mechanical theory of granular matter (and indeed of other amorphous solids) is the absence of a unique equilibrium reference state. This is of course true also for classical fluids, making the definition of ``deformation" quite vague. In contradistinction to fluids, granular matter can support shear stresses, even in the presence of significant particles flow, suggesting that an instantaneous reference state exists \cite{jiang2007brief}.
On the other hand granular matter differs from classical solids, since it responds to external loads with quasi-localized plastic events, constantly changing its grainy configuration to release mechanical stresses. It is widely accepted that understanding the spatial and temporal evolution of such plastic events is at the heart of any future granular matter theory. Consequently, intense theoretical activity attempted to develop predictive tools for the positions and magnitudes of emerging plastic events \cite{10KLLP}. Successful predictors included \emph{softness} and {thermal energy}, with the former being a geometry-based predictor, whereas the latter is based on nonlinear particle interactions \cite{schoenholz2016structural,zylberg2017local}.

Despite the success in predicting localized plastic events and their statistics, a coarse-grained continuum theory faithful to jammed granular phenomenology is still missing. 
The exception is when a 
granular matter is subjected to high pressures, in which case it 
obeys the theory of elasticity.  Upon reducing the imposed pressure and approaching the unjamming point elasticity theory breaks down.
This is most transparent when studying the density of states of vibrational modes in granular matter \cite{o2003jamming}. Indeed, the density of state of jammed granular matter indicated the emergence of diverging length scales at low pressures. At area fractions close to unjamming this length scale is finite and comparable with system size. Correspondingly, the existence of inherent length violates elasticity theory, which is scale independent. Similar observations for diverging length-scales at the onset of unjamming provide additional indications for the failure of elasticity theory \cite{hexner2018two}.

One way in which linear elasticity can fail is due to nonlinear responses. In the classical approach nonlinear moduli can be used to relate large deformations to higher order stress responses \cite{01CH}. 
The applicability of this approach to {\em amorphous solids} had been however questioned. Firstly, it was shown that in amorphous solids plastic responses appear instantaneously for any amount of strain \cite{10KLP}. Secondly, it was found that nonlinear elastic constants are not well defined in amorphous solids, having unbounded sample to sample fluctuations in the thermodynamic limit \cite{11HKLP}. Accordingly, it becomes necessary to re-examine the applicability of elasticity theory to amorphous solids.

\begin{figure*}
	\centering
	\includegraphics[width=\linewidth]{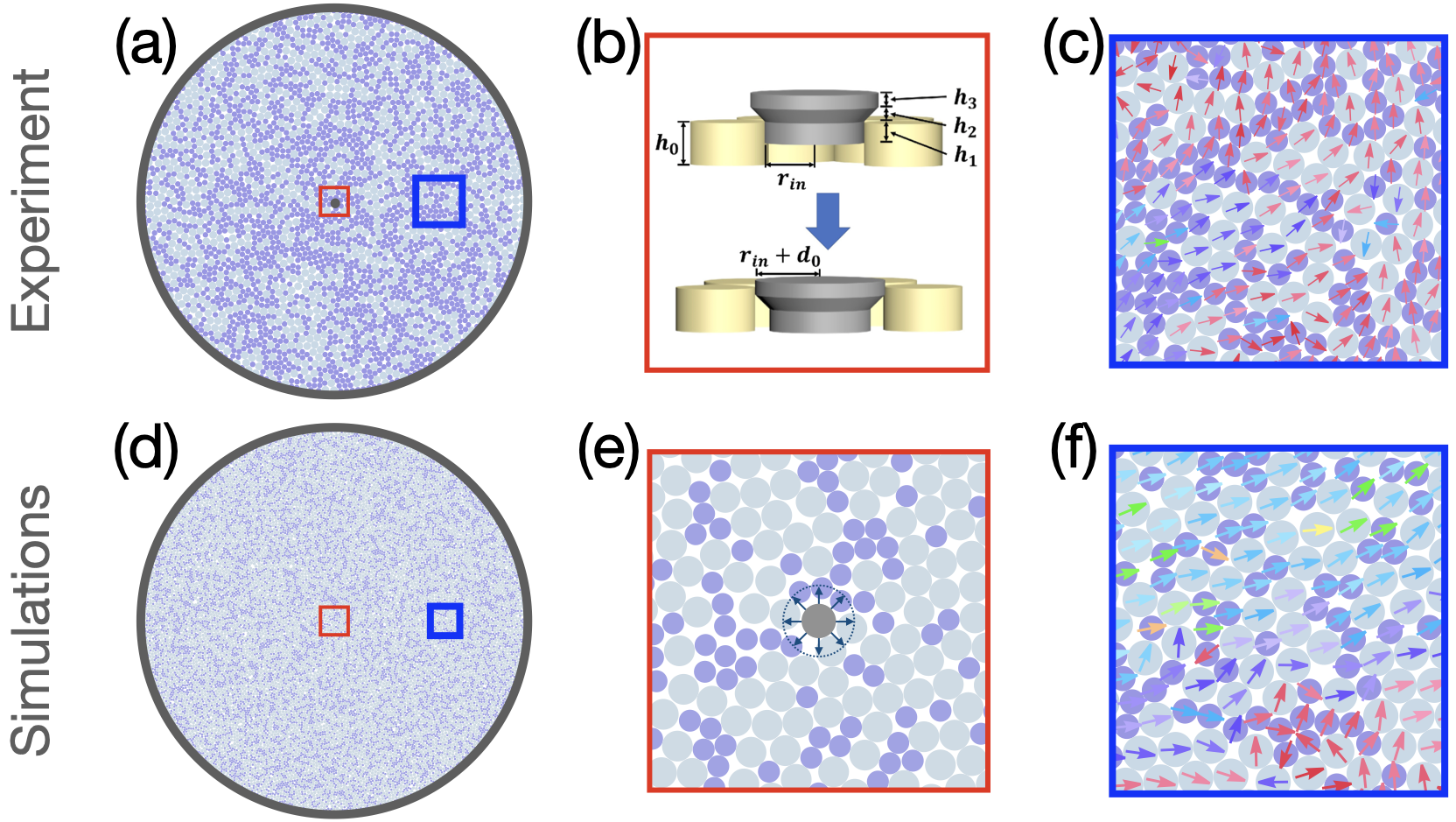}
	\caption{Experimental (top panel) and simulations (bottom panel) of the single particle expansion protocol. \textbf{Top panel}: 
		(a) Top view of the experimental system at mechanical equilibrium prior to particle expansion. The big gray circle marks the position of the circular boundary. The outer particles  
		layer consists of photo-elastic disks, which are used as pressure sensors. The bulk is filled with bi-dispersed ABS disks, see text for details. The dark dot in the center represents the conical shaped pusher used to achieve the inflation, with the detailed structure depicted in (b).
		(c) Zoom-in to the blue frame in (a). The arrows at each particle represent the resulting displacement field after inflation.
		\textbf{Bottom panel}: (d)  Top view of the numerical ground-state configuration prior to particle's inflation. (e) Single particle expansion protocol. (f) Zoom-in to the blue frame in (d). The arrows at each particle represent the resulting displacement field after inflation, colored by displacement magnitudes in arbitrary units.
		%
	}
	\label{Setup}
\end{figure*}

A recent work suggested to model granular matter as a continuum elastic medium supplemented by a distribution of responsive plastic events modeled as screening quadrupolar elastic charges \cite{21LMMPRS}. Topological conservation laws prohibit the nucleation of isolated dipolar or monopolar charges, in agreement with the identification of Eshelby-like localized events as the basic plastic degrees of freedom\cite{99ML,06ML}. 
The theory allows for two main classes of screening: (i) Weak screening where quadrupolar charges are sparse and material elements are well defined by an average quadrupole, and (ii) Strong screening where induced quadrupoles are dense, and their gradients are necessary to describe local irreversible deformations; such gradients of the quadrupolar field are identified as induced dipole.


In the weak screening regime, when the density of the induced quadrupoles is low, they act only to renormalize the elastic moduli, but they do not change the form of the theory. Linear elasticity theory can still be used to predict, for example, displacement fields that result from small strain perturbations. 
On the other hand, in the strong screening regime when gradients of the density of quadrupoles cannot be ignored, they change the structure of the theory by breaking translational symmetry. A new length scale is introduced, and mechanical responses deviate qualitatively from either linear or nonlinear elasticity, thus termed anomalous elasticity.


A prototypical deformation mode studied within this theory is the mechanical response of a granular system to the isotropic expansion (inflation) of a single granule. Due to polar symmetry, the predicted mechanical response appears independent of elastic moduli, depending only on the geometric properties of the system. In particular, the solution is universal with key features, e.g. a sign-preserving and monotonic radial displacement field. 
A moduli independent  displacement field is important when studying the breakdown of elasticity since any deviation from it cannot be reconciled by adjusting elastic moduli.  
The anomalous screening model, on the other hand, predicts a non-monotonic displacement field that depends on an inherent length-scale. 
In a recent numerical work on this problem with 2d granular model, non-monotonous displacement fields were observed, in agreement with the screening theory \cite{21LMPRWZ}.

In this paper we further investigate the theory of anomalous elasticity and screening, predicting stronger deviations from classical elasticity. In particular, we predict that the isotropically expanding grain may result with spatially oscillating and sign changing displacement fields. 
Importantly, we experimentally study below assemblies of binary frictional disks of two different sizes (to achieve an amorphous material), bounded by a circular wall.
In both experiments and simulations we equilibrate the systems at a chosen target pressure, and then inflate one disk at the center of coordinates to study the ensuing displacement field.
This work has two main results: First, we confront the suggested theory with experimental measurements and provide a strong support to the new theory. Second, we discover a new phenomenon: particles in a granular solid move inwards, in response to a radial pusher expanding at the center. This counter-intuitive result was accurately predicted by the anomalous screening theory.

For clarity and completeness, we start by briefly reviewing mechanical screening theory for a granular system in 2-dimensions, and study its interesting predictions for the single-grain expansion protocol. Then we describe experimental and numerical results and compare them to the theoretical predictions. Finally we discuss the new routes opened by the proposed theory and its potential extensions to address open problems in granular and glassy matter.

\section{Theory} In linear elasticity the equation satisfied by
the displacement field reads
\begin{equation}
	\Delta \mathbf{d} + \lambda \nabla  \left(\nabla \cdot \mathbf{d}\right)= 0 \ , \quad \lambda\equiv \frac{1+\nu}{1-\nu} \ ,
	\label{elastic}
\end{equation}
with the appropriate boundary conditions, where $\nu$ is the 2-dimensional Poisson ratio. 

\begin{figure}
	\centering
	\includegraphics[width=0.95\linewidth]{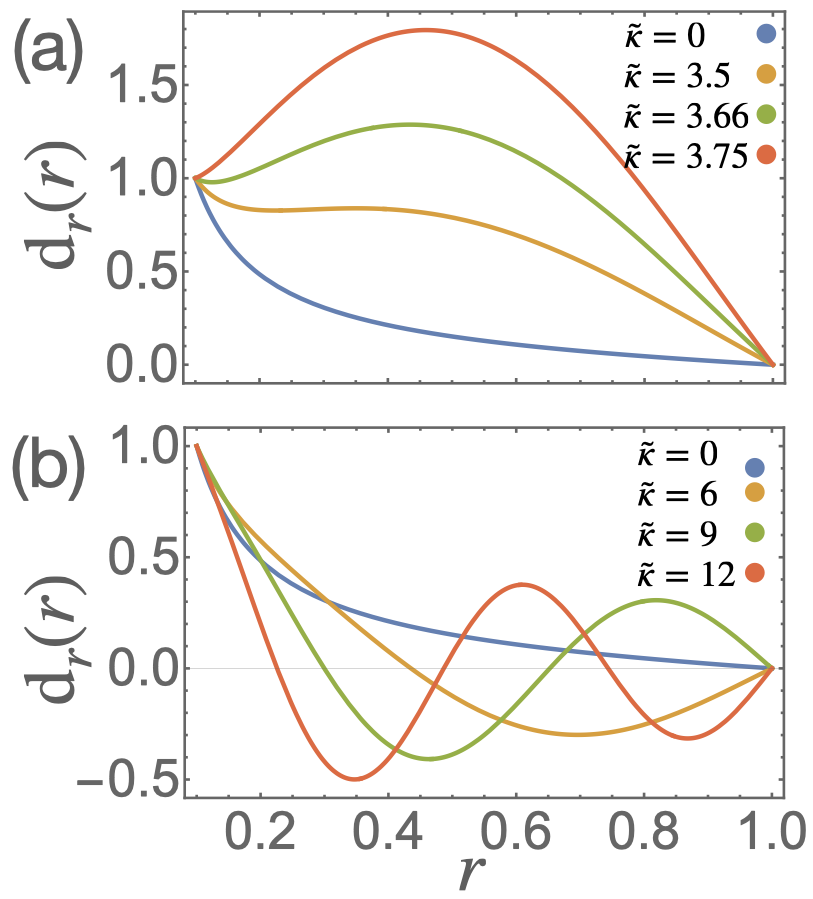}
	\caption{The solution \eqref{amazing} for various values of the screening parameter $\tilde{\kappa}$.  (a) Small values of $\tilde\kappa$, leading to weekly anomalous elastic response with the displacement field being possibly non-monotonic, without violating its sign-preserving property. 
		(b) Large values of $\tilde{\kappa}$ leading to strong anomalous elasticity with a range of non-monotonic and sign-changing  displacement field. The number of nodes in the displacement field increases with $\tilde{\kappa}$.}  
\label{Prediction}
\end{figure}

The mechanical test we are interested in is illustrated in \figref{Setup}.
We consider a circular container of radius $\rout$ containing frictional binary granular matter (\figref{Setup}(a),(d)) made of photo-elastic disks.
A disk of radius $\rin$ is located at the center, and by pulling it down it expands to a new radius $\rin+d_0$ (\figref{Setup}(b),(e)). This setup corresponds to an imposed radial displacement $\mathbf{d}(\rin) = d_0 \hat{r}$. Consequently, particles move outwards until reaching a new mechanical equilibrium (\figref{Setup}(c),(f)).
%
The polar symmetry of the problem implies that $\mathbf{d}(r) =d_r(r) \hat{r}$, in which case the equilibrium equation reduces to 
\begin{equation}
	\Delta {\mathbf{d}}=0 \ .
\end{equation}
The solution to this differential equation that satisfies the boundary conditions is
\begin{equation}
	{d}_r(r)=d_0 \frac{r^2 - \rout^2}{\rin^2 - \rout^2}\frac{\rin}{r} \ .
	\label{renelas}
\end{equation}
Upon defining dimensionless quantities $\tilde{r} = r/ \rout$, $\tilde{r}_\text{in} = \rin/\rout$ and $ \tilde{d}_r = d_r/d_0$, this solution takes a universal form
\begin{equation}
	\tilde{d}_r(r)= \frac{\tilde{r}^2 - 1}{\tilde{r}_\text{in}^2 - 1}\frac{\tilde{r}_\text{in}}{\tilde{r}} \ .
	\label{dlrenelas}
\end{equation} 
Importantly, this solution is independent of elastic moduli, and therefore deviations from this form cannot be compensated by redefinition of mechanical properties.
With the imposed displacement pushing outward, the solution (\ref{dlrenelas}) is monotonically decreasing and positive, as shown by the blue line in \figref{Prediction}(a). 

When plastic events can form to screen the elastic field, the mechanical energy density in the system is \cite{12DHP,13DHP}
\begin{eqnarray}
	W = \frac{1}{2} A^{\alpha\beta\gamma\delta} (u_{\alpha\beta} - Q_{\alpha\beta}) (u_{\gamma\delta} - Q_{\gamma\delta}) + \mathcal{F}(Q,\nabla Q) \;.
\label{eq:Energy}
\end{eqnarray}
Here $Q$ is the quadrupoles field that modifies the rest state of the system and therefore release strains as depicted by the first term in \eqref{eq:Energy}.
The function $\mathcal{F}(Q,\nabla Q)$ is a cost function penalizing for quadrupoles nucleation. 
Two main classes of screening models penalizes the nucleation of quadrupoles and their divergence \cite{21LMMPRS}: 
\begin{eqnarray}
	\mathcal{F} = \begin{cases} 
		\frac{1}{2} \Lambda_{\alpha\beta\gamma\delta} Q^{\alpha\beta}Q^{\gamma\delta} & \text{Quasi-Elastic} \\
		\frac{1}{2} \Gamma_{\alpha\beta} (\nabla_{\mu}Q^{\alpha\mu}) (\nabla_{\nu} Q^{\beta\nu}) & \text{Anomalous}
	\end{cases}\;.
\end{eqnarray}
The multipole expansion of geometric charges implies that the divergence of a quadrupole field is effectively acting as a dipole field \cite{moshe2015elastic}, hence the anomalous screening is equivalent to (totally neutral) dipole screening. 
In the quasi-elastic case, induced plastic events are sparse, described by low density of screening quadrupoles. 
In this regime the form of \eqref{elastic} remain intact, only with renormalized elastic moduli \cite{21LMMPRS}. 
%
%
On the other hand, when the density of plastic events becomes large and the gradients of their density cannot be neglected, Eq.~(\ref{elastic}) changes and assumes the form \cite{21LMMPRS}:
\begin{equation}
	\Delta \mathbf{d} + \lambda \nabla \left(\nabla\cdot \mathbf{d}\right) = -\mu \mathbf{d}
	\label{anomalous}
\end{equation}
The parameter $\mu$ is related with the nucleation energy of a 
local quadrupole gradient
and has the dimensions of length$^{-2}$. The screening effect is negligible when $\mu \, L^2 \ll 1$ where $L$ is the system size. 
Unlike the quadrupole screening, dipole screening leads to a qualitatively new behavior.  In polar coordinates the equation for the displacement field assumes the form of Bessel equation
\begin{equation}
	d_r'' +\frac{1}{r} d_r' +(\kappa^2 -\frac{1}{r^2})d_r=0 \,
\end{equation}
with $\kappa^2 \equiv \mu/(1+\lambda) $. A solution 
of this equation satisfying $d_r(r_{\rm in})=d_0$, $d_r(r_{\rm out})=0$ reads
\begin{equation}
	d_r(r)  = d_0 \frac{ Y_1(r \, \kappa ) J_1(r_\text{out} \kappa )-J_1(r \, \kappa ) Y_1(r_\text{out} \kappa )}{Y_1( r_\text{in} \kappa ) J_1(r_\text{out} \kappa )-J_1(r_\text{in} \kappa ) Y_1(r_\text{out} \kappa )} \ .
\end{equation}
As before, upon measuring $r,\rin,\kappa$ in units of $\rout$ and $d_r$ in units of $d_0$ we find
\begin{equation}
	\tilde{d}_r(r)  =  \frac{ Y_1(\tilde{r} \, \tilde{\kappa} ) J_1(\tilde{ \kappa} )-J_1(\tilde{r} \, \tilde{\kappa} ) Y_1(\tilde{ \kappa} )}{Y_1( \tilde{r}_\text{in} \tilde{\kappa} ) J_1(\tilde{ \kappa} )-J_1(\tilde{r}_\text{in} \tilde{\kappa} ) Y_1(\tilde{ \kappa} )} \ .
	\label{amazing}
\end{equation}
Here $J_1$ and $Y_1$ are the Bessel functions of the first and second kind respectively.
It should be stressed that at this point we do not have a-priori theory for the numerical values of $\tilde{\kappa}$, and in comparisons to experiments and simulations we need to fit this parameter. In the analysis of the experimental results below the values of the other parameters, namely $r_{\rm in}$ and $r_{\rm out}$ are not
fitted; they are directly taken from the data of the experiments.

\begin{figure*}
	\centering
	\includegraphics[width=\linewidth]{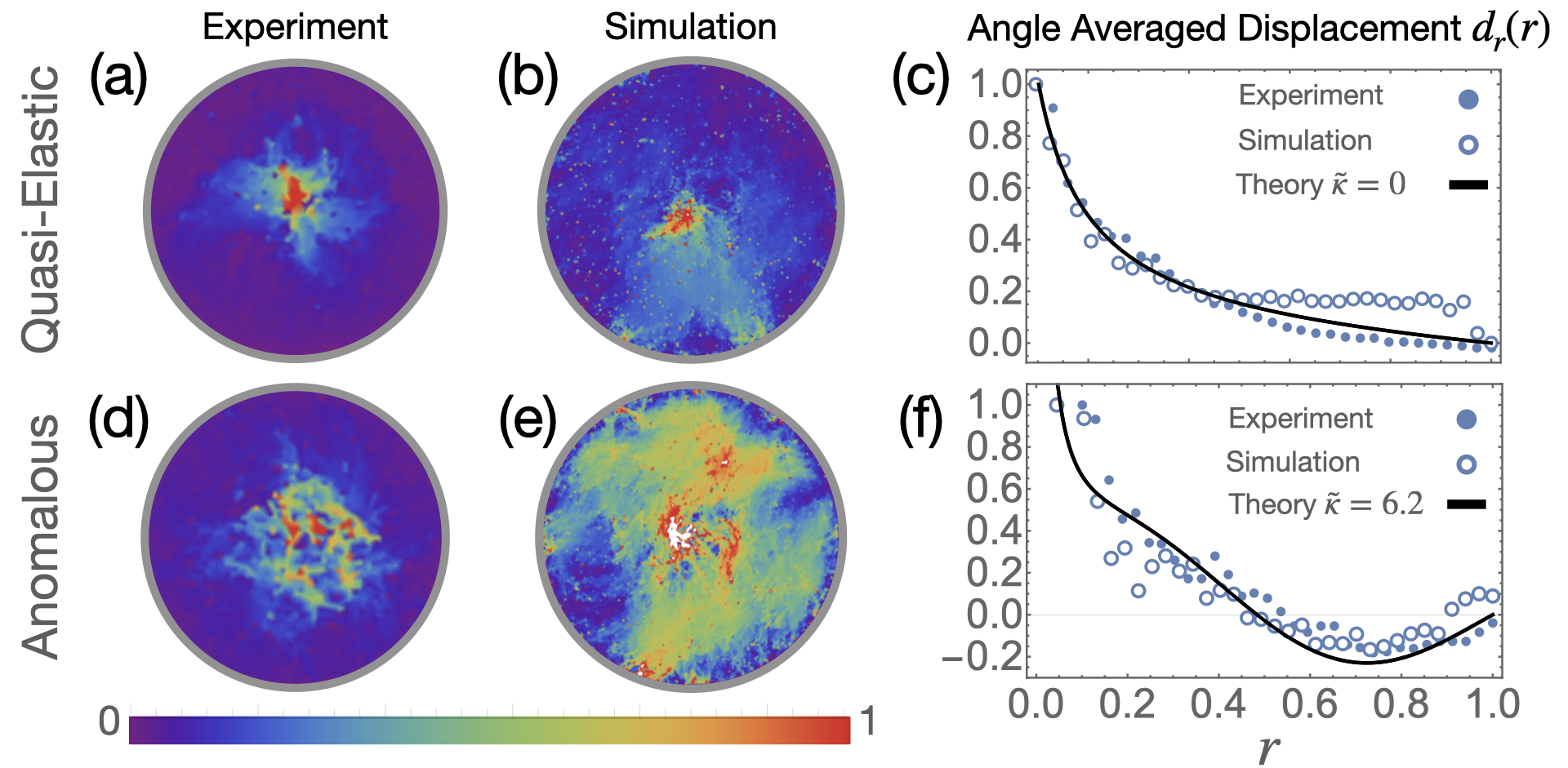}
	\caption{Mechanical response to single particle expansion with weak screening (top panel) and strong screening (bottom panel), corresponding to high pressure quasi-elastic and low pressure anomalous responses. \textbf{Top}: Experimental (a) and numerical (b) density plots of the displacement field magnitude induced by a particle expansion at the center, normalized  according to \eqref{amazing}. The two realizations are subjected to large dimensionless pressures $\tilde{P}_\mathrm{exp} \approx 10^{-3}$ and $\tilde{P}_\mathrm{sim} \approx 2.25 \times 10^{-6}$ correspondingly. (c) The angle-averaged and normalized radial displacement measured experimentally in (a) (solid dots) and numerically in (b) (open circles), compared with the quasi-linear prediction of \eqref{dlrenelas} (black solid line). 
		\textbf{Bottom}: (d,e) Density plots of the displacement field magnitude induced by a particle expansion at the center, normalized by the imposed expansion, as observed in experiment (d) and in simulations (e), when subjected to dimensionless pressures $\tilde{P}_\mathrm{exp} \approx 10^{-6}$ and $\tilde{P}_\mathrm{sim} \approx 2 \times 10^{-7}$ correspondingly. (f) The angle-averaged and normalized radial displacement measured experimentally in (d) (solid dots) and numerically in (e) (open circles), plotted together with the anomalous prediction of \eqref{amazing} (black solid line). 
	}
	\label{Results}
\end{figure*}

{\bf The set up of experiments and simulations:} The experimental apparatus consists of a circular frame with a radius of 350 mm and  height of 8 mm, that is stacked on a horizontal glass plate, 
which is shown as the blue circle in Fig.~\ref{Setup} panel (a,b). We have performed two sets of experiments using bi-disperse disks of two different stiffness constants, 
in order to produce two-dimensional amorphous packings at low and high dimensionless pressure values.

In the low dimensionless pressure experiment, we use bi-disperse acrylonitrile butadiene styrene (ABS) disks with a 1:1 number ratio, to prepare the amorphous packing. Firstly, we pad the 
outer circumference with a single layer of bi-disperse photo-elastic disks to act as pressure sensors, forming the outer layer in Fig.~\ref{Setup}(a). Secondly, 
we fill up the rest of the circular frame with the ABS disk. The radii of the large and small disks are 7.0 mm and 5.5 mm, respectively, while the 
radii of the large and small photo-elastic disks are 7.0 mm and 5.0 mm, respectively. The area fraction of the whole system is 0.836 before the inflation at the center of the system.

For the high dimensionless pressure experiment, we fill the rest of the circular frame with the same bi-disperse photo-elastic disks as the ones at the circumference. 
The number ratio of small to large disks is 2:1, and the area fraction is 0.835 before the inflation. In the experiments, we apply photo-elastic techniques to measure the forces at the circumference and then use the normal force components of the contact force between boundary and the inside disks to obtain the values of pressure. 
When all the disks are photo-elastic, we can also measure pressure from the bulk since the pressure is the same in the bulk and at the boundary.

A conically shaped pusher (cf. Fig.~\ref{Setup} (b)) is placed in the center of the system to apply the inflation there, as shown by the dark dot in Fig.~\ref{Setup} (a). The  
geometry and parameters are specified in Fig.~\ref{Setup} (b). Initially, this conically shaped pusher can contact the neighboring disks only by its skinny lower part. By pressing this pusher down smoothly, its broad cap starts to contact its surrounding disks, producing an effective inflation 
of the center disk \cite{14CSD}.

We produce an initial amorphous configuration of a desired pressure by gradually adjusting the packing fraction, while at the same time we apply random tapping to 
eliminate any potential stress and local inhomogeneity of the amorphous assembly. The images of disks configurations are captured by a $2\times 2$ 
array of high-resolution cameras above the system, whose four images can be stitched together through the calibration, using a checker board to achieve a 
spatial resolution of the disk position up to 10 pixel/mm. We can identify the position of the disks before and after the inflation at the center, 
to measure the displacement field. Specifically we report below how
the radial component of the displacement field, averaged over a circle of radius $r$, decays with the distance $r$ from the inflating screw.

Since the ABS disks have a bulk modulus of $B = 2.2$ GPa and thickness of $h_0\sim 1$ cm, it gives a microscopic pressure per disk of 
$P_{mic} = Bh_0 = 2.2 \times 10^7$ N/m. This gives, for a typical value of the desired pressure $P\sim 10$ N/m, a dimensionless pressure 
$\tilde P = P/P_{mic}\sim 10^{-6}$ in the amorphous packings of ABS disks. Similarly, the photo-elastic disks have a bulk modulus of 
$B = 4$ MPa and thickness of $h_0 = 6$ mm, which gives a microscopic pressure per disk of $P_{mic} = B h_0 = 24 \times 10^3$ N/m. 
This gives a dimensionless pressure $\tilde P = P/P_{mic}\sim 10^{-3}$ in the amorphous packings of photo-elastic disks. Additional details on this renormalization procedure are offered in the Supplementary Material. 

The simulations employ $N=16000$ disks of two radii, half with a radius $R_1 = 0.35$ and the other half with a radius $R_2 = 0.49$. All the simulation units are 
quoted in SI units. The forces between the disk act only upon contact. The contact forces, which include both normal and tangential components due to friction, are modeled according to the discrete element method developed by Cundall and Strack \cite{79CS}, combining a Hertzian normal force and a tangential Mindlin component. The exact form of these forces is presented for example in the section "Materials and Methods" in Ref.~\cite{21LMPR}. The stiffness constant in the Hertzian force law is $k_n=2\times 10^6$ N/m.
Open source codes, LAMMPS \cite{95Pli} and LIGGGHTS \cite{12KGHAP} are used to perform the simulations.
Initially the disks are placed in a circular box of fixed boundaries
with an initial area fraction $\phi=0.45$. Next, isotropic compression is implemented in a step-wise fashion by inflating each disk by a small factor (1.00004) followed by subsequent relaxation until the forces and torques on each disk are smaller than $10^{-7}$ in SI units. This process is continued until mechanically stable configurations are generated at a desired pressure $P$. The dimensionless pressure $\tilde P$ is obtained by dividing by the stiffness constant, $\tilde P =P/k_n$. 

\begin{figure*}
	\centering
	\includegraphics[width=\linewidth]{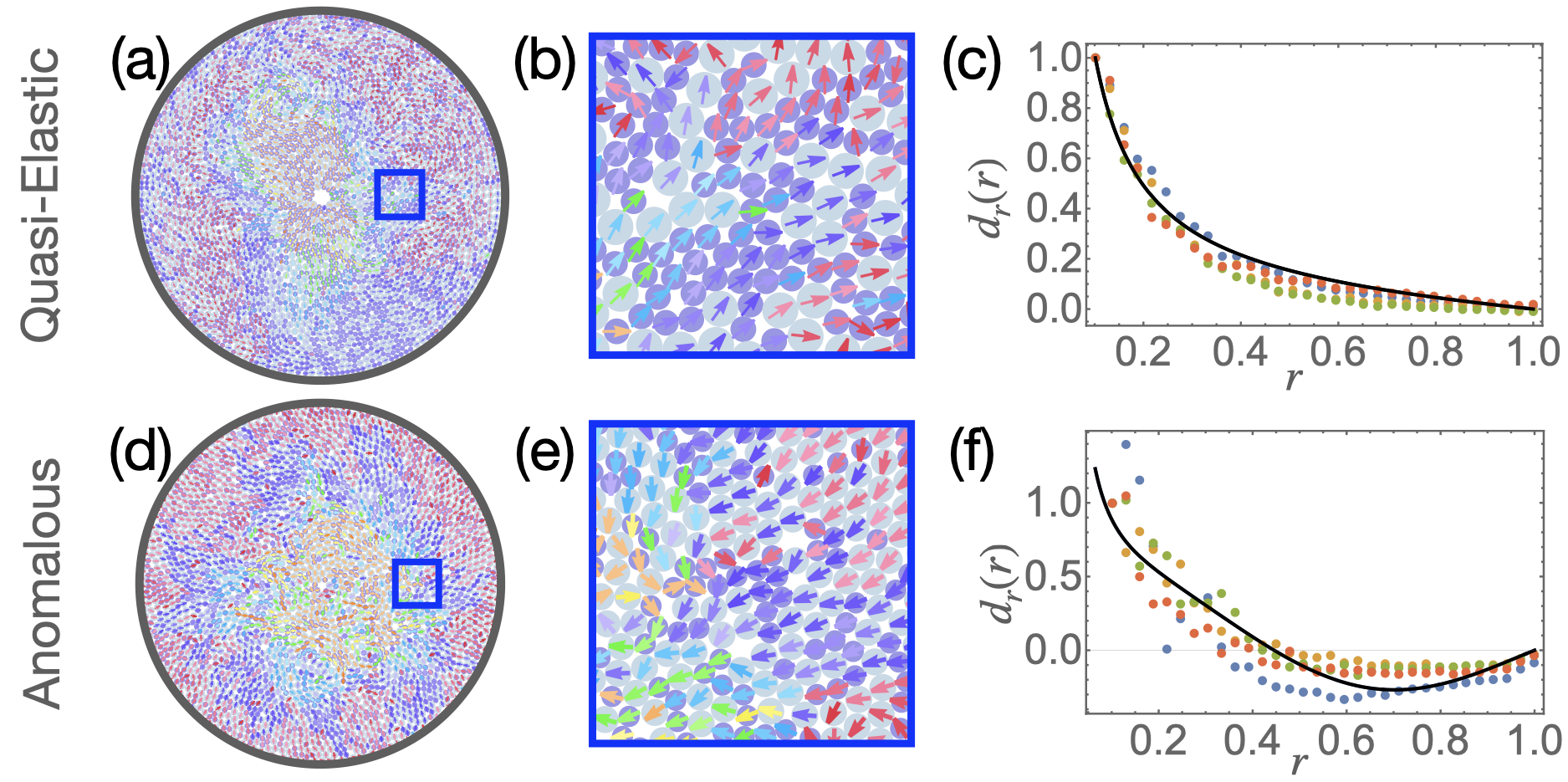}
	\caption{Experimental verification of Elastic (top) and Anomalous (bottom) mechanical responses to a single particle expansion. (a) The displacement field induced by single particle expansion in a high pressure realization, colored by their magnitudes. (b) Zoom in to the blue frame, indication the out-flow of particles, as expected from normal elasticity. (c) Multiple high-pressure realizations of angle-averaged radial displacement field, collapsing to the universal form of \eqref{dlrenelas}.
	 (d) The displacement field induced by single particle expansion in a low pressure realization, colored by their magnitudes. (e) Zoom in to the blue frame, indicating the in-flow of particles, as predicted by anomalous elasticity in \figref{Prediction}(b). (f) Multiple low-pressure realizations of angle-averaged radial displacement field, collapsing to the universal form of \eqref{amazing} with similar $\kappa$'s.
 } 
	\label{Zoomin}
\end{figure*}

After achieving a mechanically stable configuration at a
target pressure, we choose the disk with larger diameter that is
closest to the center of the simulation box and inflate it by a desired amount,
and let the system evolve until equilibrium is reached. Once we
reach the equilibrium state, we examine the displacement field
that is induced by this inflation. Specifically we study how
the radial component of the displacement field, averaged over a circle of radius $r$, decays with the distance $r$ from the inflated disk.

{\bf Results:} As stated, at large pressures we observe a paucity of plastic responses, and accordingly the radial component of the displacement field is expected to follow normal elasticity theory.
This expectation is realized in both experiments and simulations as can be seen in top-panel of  Fig.~\ref{Results}. 
%
%
Top panel shows results for high dimensionless pressures in experiments ($\tilde P = 10^{-3}$) and simulations ($\tilde P=2.25\times 10^{-6}$), introducing quasi-elastic response. 
In panels (a) and (b) we see 
experimental and numerical density plots of the displacement field, indicating the dilute distribution of induced quadrupoles.  
In panel (c) we plot the angle-average radial displacement fields corresponding to (a,b) from experiment and simulation, that are in close agreement with Eq.~(\ref{dlrenelas}).


On the other hand, at small pressures the density of plastic responses increases sharply. 
This is shown in bottom panel of Fig.~\ref{Results}, where the displacement density plots in (d,e) indicate the prevalence of high density plastic responses. 
The low dimensionless pressures in experiments is $\tilde P= 10^{-6}$ and in simulations $\tilde P = 2 \times 10^{-7}$.
To check that the displacement field is indeed plastic, we can deflate the central disk and examine whether the displacement fields annuls or remains unchanged. In the Supplementary Material we show that the latter case is observed.
In panel (f) we plot the angle-averaged radial displacement fields corresponding to (d,e) from experiment and simulation, and compared with Eq.~(\ref{amazing}), with only one fit parameter $\tilde{\kappa}$. 
We observe, for the first time, an anomalous response with 
the displacement field now becomes {\em negative} before it returns to zero at the outer boundary. The plot in Fig.~\ref{Results}(f) validates a good agreement between theory, experiment, and numeric simulations. 

The fit for both experiment and simulation shown in , is $\tilde{\kappa} = 6.2$. We emphasize that the inner and outer radii $r_\text{in}$ and $r_\text{out}$ are not fitted, but rather taken directly from the experimental and numeric data. 

To test  robustness of the mechanical response we next perform multiple experiments of different realizations subjected to similar dimensionless pressures. In \figref{Zoomin} top-panel we show results for high dimensionless pressure ($\tilde P = 10^{-3}$) with quasi-elastic response. In (a) we show the displacement field colored by its magnitude, (b) Zoom-in to a portion of the material showing an outwards flow, and (c) Angle-averaged radial displacement fields compared with the universal form \eqref{dlrenelas}.
 In \figref{Zoomin} bottom-panel we show results for low dimensionless pressure ($\tilde P = 10^{-6}$) with anomalous response. In (d) we show the displacement field colored by its magnitude, (e) Zoom-in to a portion of the material showing an {\em inwards} flow, and (f) Angle-averaged radial displacement fields compared with \eqref{amazing} with $\kappa = 6.2$.

{\bf Discussion}: It is quite obvious that the radial displacement fields sown in \figref{Results}(f) and \figref{Zoomin}(f)
cannot be possibly assigned to standard linear elasticity theory. We note that the present experiment, as well as the experiment reported in Ref.~\cite{14CSD}, indicate strongly that elasticity theory needs to be revisited in the context of amorphous solids. The appearance of a range of distances in which the displacement is negative, is directly related to plastic responses that reduce the pressure in the bulk, resulting in disks displaced inwards, even though the inflation at the origin points outwards. The fit of the data to the theoretical prediction Eq.~(\ref{amazing}) is quite remarkable, especially since there is only one available fit parameter $\tilde{\kappa}$. 
We note that the theoretical result of Eq.~(\ref{amazing}) predicts that for larger values of the screening parameter $\tilde{\kappa}$, spatial oscillations will form in the displacement field. A hint for that is observed in the numerical simulations as shown in panels (f) of Fig.~\ref{Results}, where the displacement field has a short range of positive displacement close to the outer boundary, which decays to zero from above.
It should be stated however that the fits with numeric simulations in these figures are not perfect, and whether this is an indication of the failure of the continuum theory on scales of the disks, or whether this is due to the a spatial dependence of $\tilde{\kappa}$, is still unknown, and further research is required. 

The change from quasi-elastic behavior at high pressure, cf. top-panel of Fig.~\ref{Results}, to anomalous behavior at low pressure, cf. bottom-panel of Fig.~\ref{Results}, calls for further study. Presently it is not clear whether this change is gradual or sharp, with the second option being an indication of a possible phase transition at some intermediate value of the pressure, separating the two types of behavior.
At present our systems, both in simulations and experiments, are too small to
support a sharp phase transition. We find realizations at low pressure that
show quasi-elastic behavior, and vice versa, realizations at high pressure that show displacement fields that cannot be fitted to the quasi-elastic solution. We thus defer the study of the possible phase transition to future work where much more data and other types of amorphous solids will be considered. It is noteworthy however that the type of transition alluded to, if it exists, is very reminiscent of the hexatic phase transition in 2-dimensional melting \cite{78HN,79You}, where a screening by dipoles replaces low-density quadrupoles. This analogy will be examined more closely in future publications. 

{\bf acknowledgments}: This work had been supported in part by the Israel Science Foundation (collaboration with China) and the
Minerva Center for ``Aging, from physical materials to human tissues" at the Weizmann Institute.  MM acknowledges support from the Israel Science Foundation (grant No. 1441/19). S.R. acknowledges the support of the Science and Engineering Research Board, DST, India under grant no. SRG/2020/001943 and the IIT Ropar under ISIRD grant. J.S and J.Z are supported by the the National Natural Science Foundation of China (NSFC) under (No. 11774221 and No. 11974238) and 
also by the Innovation Program of Shanghai Municipal Education Commission under No. 2021-01-07-00-02-E00138.

\subsection*{Supporting Information (SI)}

TO BE EDITED IN A SEPARATE  FILE.

$r_\text{in}=7.0$ mm, $d_0=0.7$ mm, $h_1=2.0$ mm, $h_2=0.7$ mm, and $h_3=3.0$ mm for the conical shaped pusher in the low dimensionless pressure experiment, and $r_\text{in}=14.0$ mm, $d_0=1.4$ mm, $h_1=2.0$ mm, $h_2=1.4$ mm, and $h_3=3.0$ mm in the high dimensionless pressure experiment. The value of $h_0$ depends on the pusher type: $h_0=8.0$ mm for the ABS disks, and $h_0=6.0$ mm for the photo-elastic disks. 

\bibliography{friction.anomalous}

\end{document}